\documentclass[prb,preprint]{revtex4} 
\input{epsf}  
\usepackage{graphicx} 
 
\begin{document}  

\title{An ``All Possible Steps'' Approach to the Accelerated Use of Gillespie's Algorithm}  

\author{Azi Lipshtat}  

\affiliation{  
Department of Pharmacology and Biological Chemistry, 
Mount Sinai School of Medicine, 
New York, NY 10029}  
\email[E-mail: ]{Azi.Lipshtat@mssm.edu}
 
\begin{abstract}  
Many physical and biological processes are stochastic in nature.
Computational models and simulations of such processes are a mathematical and computational challenge.
The basic stochastic simulation algorithm was published by D. Gillespie about three decades ago 
[D.T. Gillespie, J. Phys. Chem. {\bf 81}, 2340, (1977)]. 
Since then, intensive work has been done  to make the algorithm 
more efficient in terms of running time. All accelerated versions of the algorithm
are aimed at minimizing the running time required to produce a stochastic trajectory in state space.
In these simulations, a necessary condition for reliable statistics is averaging over a large number
of simulations. 
In this study I present a new accelerating approach which does not alter the stochastic algorithm, but 
reduces the number of required runs. By analysis of collected data I demonstrate 
high precision levels  with fewer simulations.
Moreover, the suggested approach provides a good estimation of statistical
error, which may serve as a tool for determining the number of required runs.

\end{abstract}  
  
 
\maketitle  

\section{introduction}
\label{Sec:intro}


Statistics and dynamics of stochastic systems have attracted a great deal of interest 
in many scientific fields, including physics, ecology, chemistry and biology. 
\cite{vanKampen1981,bartlett1961,Karlin1998}. 
The small and discrete number of reactant molecules in a  cell leads
to fluctuation-dominated dynamics. This type of dynamics appears to have 
important consequences in biology \cite{McAdams1997,McAdams1999,Elowitz2002}. 
Time evolution of such systems cannot be treated by standard continuous-time 
deterministic differential equations. For proper mathematical modeling of 
fluctuating systems, a probabilistic method is often necessary.
Analytical solution of the probabilistic Master Equation is 
rarely available.
In most cases the data is obtained from numerical simulations.
There are several algorithms for simulating stochastic systems 
\cite{gillespie1976,Gillespie1977,Gibson2000,Cao2004,Gillespie2001,Haseltine2002,Gillespie2003,Rao2003,Rathinam2003}.  
However, all these methods are focused on the simulation algorithm, but not on efficient analysis of the 
output.
 A high precision level requires many runs of the algorithm. 
Here we propose a method of reducing the number of runs which are required for reliable estimation of moments.
By doing so we reduce significantly the computational resources needed for stochastic simulations.
We start with a general description of stochastic systems 
and the standard stochastic simulation algorithm.
Our new analysis method is presented in Sec. \ref{Sec:APS}.
In Sec. \ref{Sec:timecourse} we extend this approach to time course simulations.
A discussion and summary are presented in Sec. \ref{Sec:summary}.

\subsection{Stochastic systems} 

A stochastic system consists of $N\ge 1$ molecular species, $\{S_1,\ldots, S_N \}$, interacting
through $M\ge 1$ chemical reactions $\{R_1,\ldots , R_M\}$. The state of the system is defined
by the molecular populations ${\bf X}(t)=\left(X_1(t),\ldots,X_N(t)\right)$. This is a random variable whose
dynamics is determined by the reactions. For example, an occurrence of the reaction  
$S_1+S_2\rightarrow S_3$ changes the system's state from 
$\left(x_1,x_2,x_3\right)$ to $\left(x_1-1,x_2-1,x_3+1\right)$.
The stoichiometric vector of a reaction $R_\mu$ is denoted by ${\bf\nu}_\mu$. 
The $j$th component of this vector is the discrete change in the population size 
of $S_j$  resulting from a single occurrence of an $R_\mu$ reaction.
Thus, the reaction takes the system from state ${\bf x}$ to  ${\bf x}+{\bf \nu}_\mu$.
Given that the state of the system at time $t$ is ${\bf X}(t)={\bf x}$, the probability that a reaction $R_\mu$
will occur during a time interval $dt$, is given by the {\em propensity function}
$a_\mu ({\bf x})dt$. This function takes into account both the constant physical coefficients and the combinatorial factors 
which are state-dependent. Given ${\bf X}(t)={\bf x}$, the probability that  the next reaction will 
be $R_\mu$, and that 
it will occur 
in the time interval $[t+\tau,t+\tau + d\tau)$, is given by
\begin{equation}
p(\tau,\mu|t,{\bf x})=a_\mu({\bf x})\exp(-a_0({\bf x})\tau)
\label{Eq:p}
\end{equation}
\noindent
where 
\begin{equation}
a_0({\bf x})=\sum_{\mu=1}^M a_\mu({\bf x}).
\end{equation}
\noindent
The probability in Eq. \ref{Eq:p} can be interpreted as a product of two  independent probabilities: 
The probability to have $R_\mu$ as the next reaction is $a_\mu/a_0$, and the probability that next reaction
will take place after about $\tau$ time units is $a_0({\bf x})\exp(-a_0({\bf x})\tau)$.

\subsection{Gillespie's algorithm}

The standard algorithm for stochastic simulations was developed by Gillespie \cite{Gillespie1977}.
The algorithm follows a trajectory $\left({\bf X}(t_0), {\bf X}(t_1),\ldots,{\bf X}(t_{\rm final})\right)$ 
of the system in the state space. 
The actual time interval between 
two successive reactions  is a random variable distributed exponentially with average $a_0^{-1}$.
Thus, the algorithm can be summarized as follows:
\begin{eqnarray}
&&\mbox{ {\em Initialize}  $t=t_0$, {\bf X}={\bf X}($t_0$).}\\
\nonumber
&&\mbox{{\em While}  $t<t_{final}$}\\
\nonumber 
&&\hspace{1cm}\mbox{-\textit{calculate rates} $a_\mu$  \textit{and} $a_0$  \textit{ for current state}  {\bf x}}.\\
\nonumber 
&&\hspace{1cm}\mbox{-\textit{draw} $\tau$  \textit{from exponential distribution with mean} $a_0^{-1}$}.\\
\nonumber 
&&\hspace{1cm}\mbox{-\textit{advance time from} $t$ \textit{to} $t+\tau$.}\\
\nonumber 
&&\hspace{1cm}\mbox{-\textit{pick next reaction to  be} $R_\mu$ \textit{with probability} $a_\mu/a_0$.}\\
\nonumber 
&&\hspace{1cm}\mbox{-\textit{update  state} ${\bf x}$ \textit{to} ${\bf x}+\nu_\mu$.}
\label{Eq:Gillespie}
     \end{eqnarray}  

The algorithm is  computationally expensive and intensive efforts have gone into making it more efficient 
\cite{Lok2004,Gibson2000,Cao2004,McCollum2006,Gillespie2001,Haseltine2002,Gillespie2003,Rao2003,Rathinam2003}.
In his original work, Gillespie suggested two versions of the algorithm - 
the Direct method (DM) which is shown in \ref{Eq:Gillespie}  and the First Reaction method (FRM) \cite{Gillespie1977}. 
Each of these was later optimized and found to be superior in certain cases
\cite{Gibson2000,Cao2004,McCollum2006}. In addition to those exact algorithms,
there are also some approximations which significantly accelerate  the simulation performance,
with minor loss of precision \cite{Resat2001,Gillespie2001,Haseltine2002,Gillespie2003,Rao2003,Rathinam2003}.
The improvement is being done  by either hybridization (simulate some of the species in a deterministic manner) 
or  averaging over time intervals.

\subsection{Data collection from simulations}

Gillespie  writes\cite{Gillespie1977}:
\begin{quote}
``If it is desired to estimate any of the {\em moments} $X_i^{(k)}(t)$ of the grand probability function... 
then it will be necessary to make {\em several} simulation runs from time $0$  to the chosen time $t$...
Any moment  $X_i^{(k)}(t)\equiv \langle X^k_i\rangle_t$ may then be estimated directly as the average of the $k$th 
power of the numbers found for $X_i$ at time $t$ in these runs.''
\end{quote}

The first moment is of interest in most cases, but higher moments may also be important.
The second moment $ \langle X^2_i\rangle$ determines the variation,
and combined moments $ \langle X_iX_j\rangle$ give reaction rate of second order reactions. 
Even though estimation of the various moments is the typical aim of a stochastic simulation, one can hardly find 
explicit reference to the moment calculation in the literature.
Very often the algorithm is described in great detail. The  selection
procedure for  
the next reaction   and choice of data structure are elaborately explained.
However,  there is no mention of the translation of the resulting trajectory into moments. 
It is implicitly assumed that the statistics extraction process
is straightforward. 
However, there may be different ways of performing the analysis. 
If a time course of a moment is needed, one has to define a set of
time points. These are not the reaction times, which are different for 
any execution of the algorithm.   
The values of the moment at these time points are  recorded and averaged over many runs of the algorithm.
In the more common  case, the focus is on steady state values of the moments, rather than on transients.
It is assumed that there is a steady state distribution, and the probability $P({\bf x})$ to find the system in state
${\bf x}$ does not change in time. In that case, one should average not only over many runs but also over time within 
each run of the simulation. In practice, one has to calculate the sum
\begin{equation}
\sigma^k_j=\sum_i X^k_j(t_i)\tau_i,
\end{equation}        
\noindent
where $X^k_j(t_i)$ is the value of the moment $X^k_j$ at time point $t_i$ in which the $i$th step (reaction) was taken,
and $\tau_i=t_{i+1}-t_i$ is the time to the next reaction. After the simulation run is completed, 
the moment is given by
\begin{equation}
\langle X_j^k\rangle = {1\over t_{\rm final}-t_0}\sigma^k_j.
\end{equation}
\noindent
The moments can also be calculated using an indirect approach. In this approach, the stationary probability 
distribution $P({\bf x})$  
is measured during the simulation. Any moment can be deduced later from this  distribution.
Calculation of probabilities, rather than direct estimation of moments is advantageous in systems
with multiple stable states, where the average moments obscure the existence of several possible stable states
\cite{Lipshtat2006,Loinger2007}.  
The probability distribution $P({\bf x})$ is calculated as the fraction of time that the system spent at state ${\bf x}$. 
If the system remained in state  ${\bf x}$ for time $\tau$, then the non-normalized probability 
$q({\bf x})$
has to be updated:
\begin{equation}
q({\bf x})=q({\bf x})+\tau,
\end{equation}
where $q({\bf x})$ was initialized to zero at the beginning of the simulation. 
The probability distribution $P({\bf x})$ is given at the end of the simulation by
\begin{equation}
P({\bf x})={q({\bf x}) \over t_{\rm final}-t_0}.
\end{equation}
\noindent
We refer to this method of calculation as the ``Trajectory Following'' approach.

To make sure that states are sampled according to the correct distribution, one has to 
start collecting the data not immediately at the beginning of the simulation, but after some
equilibration time. During that time the system should arrive its stationary probability distribution.
Including transient in the collected data may change the calculated moments, and require much longer simulations, 
so that this change will  be negligible.   

Either in the direct or indirect methods, one has to repeat the simulation many times and average the results over a large ensemble of 
simulations. The reason is that in the typical case there are states whose  probability is low, which are visited 
during a standard run of a simulation for short durations and a very small number of times, if at all. 
Including the effect of these rare events on the 
average moment requires a large enough number of visits, so that the portion of  time in which the system is 
in these states 
would be proportional to their probabilities. 
The error in stochastic simulations decreases as $1/\sqrt{N}$, where $N$ is the number of runs \cite{Binder1997, Fishman1996}. 
Thus high precision requires
large $N$ and longer running time.

\section{The ``All Possible Steps'' approach}
\label{Sec:APS}

Here we suggest a novel approach for  calculating the probabilities $P({\bf x})$, namely the {\em ``All Possible Steps''} (APS) method. 
This method requires a smaller number of runs for a given precision, and thus makes the whole simulation more efficient.
The ``{\em All Possible Steps}''  method is based on calculating the probability of visiting 
 states  without actually  visiting them. At any step in the simulation,
the probability for occurrence of reaction $R_\mu$ and moving the system from ${\bf x}$ to ${\bf x}+\nu_\mu$ is
given by  $a_\mu({\bf x})/a_0$.
In the standard methods, we choose one of the possible reactions and update the moment or the probabilities accordingly.
In the APS approach, we update the collected data 
by considering  {\em all} possible steps, each with a weight proportional to its probability.
At the $i$th step, the system is at state ${\bf x}={\bf x}(t_i)$. There are several possibilities for the next state.
There is a probability $a_\mu({\bf x})/a_0({\bf x})$ to choose reaction $R_\mu$ and change the state of the system
to ${\bf x}+\nu_\mu$. If this happened, the system will be at that state for a duration $\tau_\mu$ which is taken from an
exponential distribution with average $\left(a_0({\bf x}+\nu_\mu)\right)^{-1}$. Thus, we consider this step as if it was done
and update the (non normalized) probability by adding $\left(a_\mu({\bf x})/a_0({\bf x})\right)\tau_\mu$.
For the purpose of statistics collection, we consider virtually {\em all} possible reactions and update {\em all} probabilities
respectively.
Then we choose one of them and update the state accordingly. To keep the ratio between probabilities correct, we use the same
random number when calculating the duration $\tau_\mu$ for each of the possible steps. When the run is over, the 
probabilities have to be normalized. Since we considered many steps simultaneously, the normalization is not done by the total
running time. One has to divide the probabilities by their sum, so that the sum of all probabilities would be equal to unity.
The differences between data collection in the standard Trajectory Following approach and 
in the APS method for steady state simulations are  summarized in Table \ref{Tab:compare_ss}.
 
\subsection{Example: Protein dimerization}

As an example we take a schematic process of   
protein dimerization, which includes three reactions: production, degradation and dimerization.
\begin{eqnarray}
\emptyset&\stackrel{k_1}{\longrightarrow}&S_1\\
\nonumber
S_1&\stackrel{k_2}{\longrightarrow}&\emptyset\\
\nonumber
S_1+S_1&\stackrel{k_3}{\longrightarrow}&S_2
\label{Example1}
\end{eqnarray}
In this example, the Master equation is solvable analytically.
In a different context, it was shown \cite{Biham2002,Green2001} that the probability to have $n$ copies
of $S_1$ is 
\begin{equation}
P_{\rm exact}(n) = { 2^{{1 \over 2} ({k_2 \over k_3} -1)} \over n!}
\left( \sqrt{k_1 \over k_3}\right)^{n}
{ {I_{k_2/k_3+n-1} \left(2\sqrt{k_1/k_3}\right)} 
\over 
{I_{k_2/k_3-1} \left(2\sqrt{2 k_1/k_3}\right)}} ,  
\label{eq:PnB}
\end{equation}
\noindent
and the average copy number is
\begin{equation}
\langle S_1 \rangle = \sqrt{k_1 \over {2k_3}} 
{ {I_{k_2/k_3} \left(2\sqrt{2 k_1/k_3}\right)} 
\over 
{I_{k_2/k_3-1} \left(2\sqrt{2 k_1/k_3}\right)} }. 
\end{equation}
\noindent 
where $I_x(y)$ is the modified Bessel function.
A state in this system is defined by the number $n$ of $S_1$ copies.
In the Trajectory Following approach, at any state $n$, only the probability $p(n)$ 
of the current state
will be updated
\begin{equation}
\label{Eq:update_p}
{q}(n) \longrightarrow {q}(n)+\tau.
\end{equation}
\noindent
However, in the APS method we consider all possible steps, even those who were not 
chosen for actual move. Hence, after the system moved to state $n$, 
we update all possible states  for the next step, 
namely $n-2,n-1$ and $n+1$, with appropriate weights:
\begin{eqnarray}
\label{Eq:update_all_p}
\nonumber
&&q(n-1) \longrightarrow q(n-1)+\tau_{n-1}a_{n\rightarrow n-1}/a_0\\
&&q(n+1) \longrightarrow q(n+1)+\tau_{n+1}a_{n\rightarrow n+1}/a_0\\
\nonumber
&&q(n-2) \longrightarrow q(n-1)+\tau_{n-1}a_{n\rightarrow n-2}/a_0,
\end{eqnarray}
\noindent
where $\tau_{n}$ is the random time duration the system would spend in state $n$, had this 
 been the next step, and $q(n)$ is the probability to be at state $n$, before normalization. 
 
Updating several probabilities at any step requires of course more 
computations. However, since these are only standard arithmetic operations,
there is no significant overhead in terms of running time. The most expensive steps in a 
stochastic simulation are the generation of pseudo random numbers and taking decisions based on those numbers. 
In the APS approach there is no need to have any extra random numbers and the update of 
probabilities is simple, so the extra running time is negligible. Furthermore, the whole analysis can
be completely decoupled from the trajectory production process.   
One can produce the trajectory using the standard algorithm, and at a later 
time make the analysis and calculate the moments using the APS approach.

\begin{figure}
	\centering
		\includegraphics{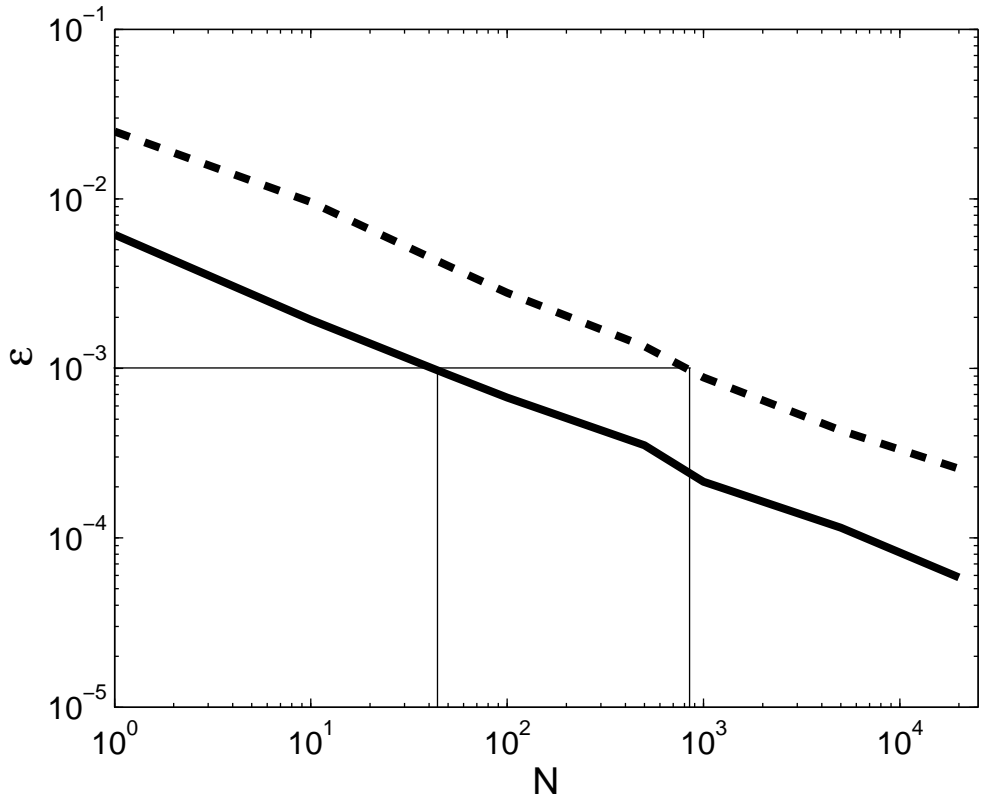}
	\caption{Error measurement (Eq. \ref{Eq:epsilon}) as function of $N$, number of runs. 
	$\varepsilon$ was calculated for the {\em same} set of simulations using both the Trajectory Following
	approach (dashed line) and the APS method (solid line).  The rates are $k_1=5,k_2=2,k_3=2$.  
	Initial condition is $S_1(0)=S_2(0)=0$. Each simulation ran for 10 time units, which is enough for attaining steady state, and then 
	statistics was collected for 150 time units ($t_0=10,t_{\rm final}=160)$). For each value of $N$ the error was averaged over an ensemble of 100 realizations of $N$ simulations. The thin lines demonstrate the 19 fold improvement in case of  $\varepsilon = 10^{-3}$. See text for details. }
	\label{Fig:dimers1}
\end{figure}

Accuracy of simulation can be measured by comparison with the exact  solution (\ref{eq:PnB}).
The error in calculation of the first moment $\langle S_1 \rangle$ is shown in Fig. \ref{Fig:dimers1}.
The relative deviation 
\begin{equation}
\varepsilon={\left|\langle S_1 \rangle_{\rm exact} - \langle S_1 \rangle_{\rm sim}  \right| \over \langle S_1 \rangle_{\rm exact}}
\label{Eq:epsilon}
\end{equation}
was calculated using both the Trajectory Following approach and the APS method. 
In both cases the error decreases as $1/\sqrt{N}$ where $N$ is the number of runs, as expected.
However, the error in the APS approach is in average about 4 times smaller  than the error in the standard Trajectory Following
method. Thus, to obtain a specific precision, one needs about 16 fold more simulations  in the standard
method than in the APS method. For example, for $\varepsilon=10^{-3}$ (precision of $0.1\%$), average over 874  simulations
is required at the standard method, whereas 47 runs  are enough in the APS approach (values obtained from interpolation of 
$N(\varepsilon)$. See thin lines in Fig. \ref{Fig:dimers1}).
This is almost a  19 fold improvement.

The distribution of errors in the APS method came out to be much narrower than that of the 
Trajectory Following approach. To show this, I ran 1000 simulation sessions. Each session 
continued until  a precision level  of $\varepsilon=10^{-3}$ was  obtained. In the APS
approach, $76\%$ of the simulations ended with an estimation for $\langle S_1 \rangle_{\rm sim}$ which is
within $\pm1\%$ of the exact  value $\langle S_1 \rangle_{\rm exact}$ (and about $9\%$ within $\pm0.1\%$), whereas in the 
Trajectory Following
method only 22$\%$ were within the $1\%$ vicinity of $\langle S_1 \rangle_{\rm exact}$ (and $2\%$ in the $0.1\%$ range).

\section{time course simulations}
\label{Sec:timecourse}

In principle, it is easy to use the same approach for time course simulations. In this case,
one should update only the probabilities at those time points which fall into the interval
$[t_i,t_i+\tau_\mu)$. The probability to find the system at these time points in state ${\bf x}+\nu_\mu$ 
is increased by $a_\mu({\bf x})/a_0$.
The normalization by the end of the simulation would be such that the sum of probabilities at each time point
will be equal to one. This procedure is summarized in Table \ref{Tab:compare_tc}.

\noindent
In practice, it turns out that for this type of simulations, the improvement of the APS method is lower than
in the steady state case. 
In time course simulations, the rare states are important because they lead to different trajectories. 
In the APS method we do not compute these hypothetical 
trajectories, and thus the resulting dynamics is not significantly 
different from that of the standard algorithm.

\begin{figure}
	\centering
		\includegraphics{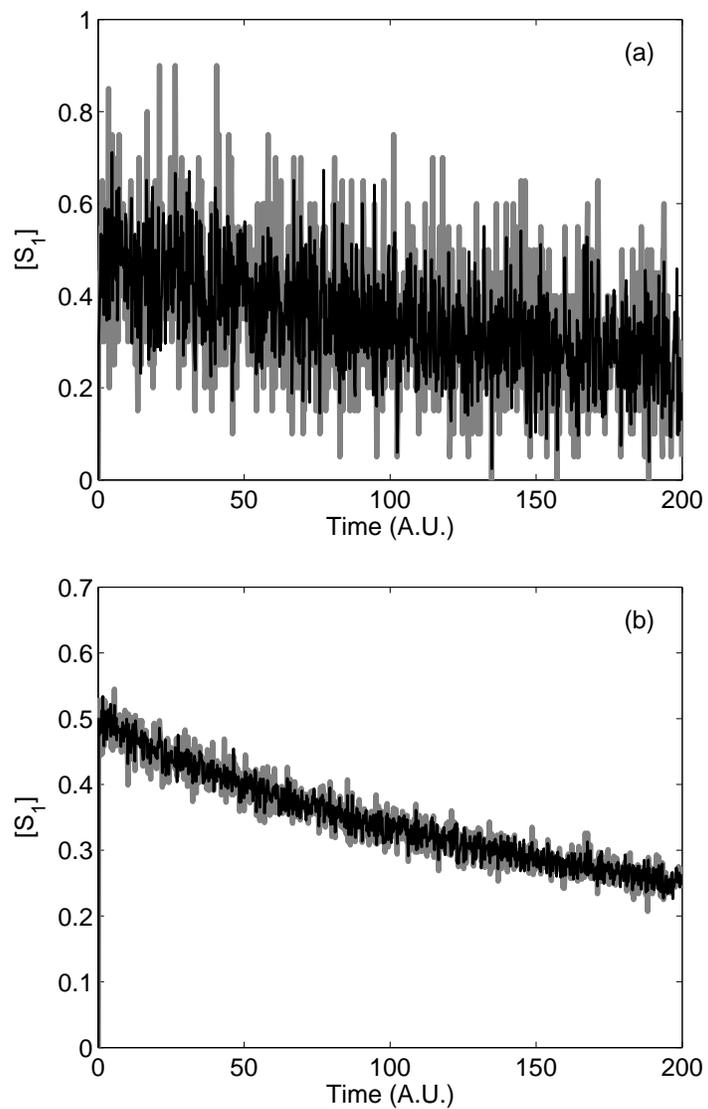}
	\caption{Dynamics of $[S_1]$  for reactions \ref{Eq:example2}. Both Trajectory Following
	approach (gray) and APS approach (black) show similar dynamics, but the Trajectory Following
	results show more fluctuations.  Averaged over (a) 20 simulations, (b) 1000 simulations.}
	\label{Fig:timecourse}
\end{figure}

Nevertheless, there are cases in which  the APS method is useful for time course simulations.  
In many systems, there is a separation between ``slow'' and ``fast'' reactions or species. In fact, some
stochastic simulation algorithms are based on this separation \cite{Rao2003,Puchalka2004,Cao2005}.
Highly reactive intermediate species, such as  enzyme-substrate complexes, usually obey the quasi-steady-state assumption
(known also as the pseudo-steady-state assumption). This means that the typical time scale for a significant change in
populations of these species is much longer than the sampling time. As a result, at any given time, these species can be 
considered to be close enough to steady state. Thus, using the APS method for simulating the stochastic system
does not affect the results of the ``slowly reacting'' species, but definitely improves the statistics as far as ``fast''
species are concerned.
We demonstrate this in a slow dimerization reaction of fast species (example ${\bf 6.2}$ of Ref. \onlinecite{Cao2005}).
The system consists of three species, out of which  two  are reactive:
\begin{eqnarray}
\nonumber
S_1&\stackrel{k_1}{\longrightarrow}& S_2,\\
\nonumber
S_1&\stackrel{k_{-1}}{\longleftarrow}& S_2,\\
S_1+S_1 &\stackrel{k_2}{\longrightarrow}& S_3,
\label{Eq:example2}
\end{eqnarray}

\noindent 
where the reaction rate coefficients are given by $(k_1,k_{-1},k_2)=(200,1,1)$ and at initial time $[S_1]=0,[S_2]=100,[S_3]=0$.
The net change in the $S_1$ population size is close to zero, and thus $S_1$ is in quasi-steady-state.
In Fig. \ref{Fig:timecourse} we present the dynamics of $[S_1]$, as obtained from
standard averaging and from using the APS method. Even though this is not a real steady state,
the change in the population size is slow.

\subsection{The virtual population approach}

The concept of considering virtual steps can be beneficial in the 
context of time course simulations in an additional way. 
In most systems there are products 
which are not involved in any reaction as reactants. $S_3$ is such a species in our last example.
Since these species are not reactive, their population size has no effect on the system dynamics.
Thus, there is no need to follow the real discrete population size. Instead, one can update the {\em virtual} 
population size at  any step of simulation, whether or not a real production reaction took place in that step. 
The update should be by the reaction rate (as obtained from the discrete system state) multiplied by the time step size.
In our example, we increased the population size of $S_3$ at any step by $k_2[S_1]\left([S_1]-1\right)\tau$, where $[S_1]$  is the discrete
population size of $S_1$ and $\tau$ is the time step, calculated by the algorithm as function of the system's state.
Mathematically, this approach is equivalent to calculating $[S_3]$ by
\begin{equation}
[S_3(t)]=[S_3(0)]+\int_0^t {d[S_3] \over dt'}{\rm dt'}.
\end{equation}

\begin{figure}
	\centering
		\includegraphics{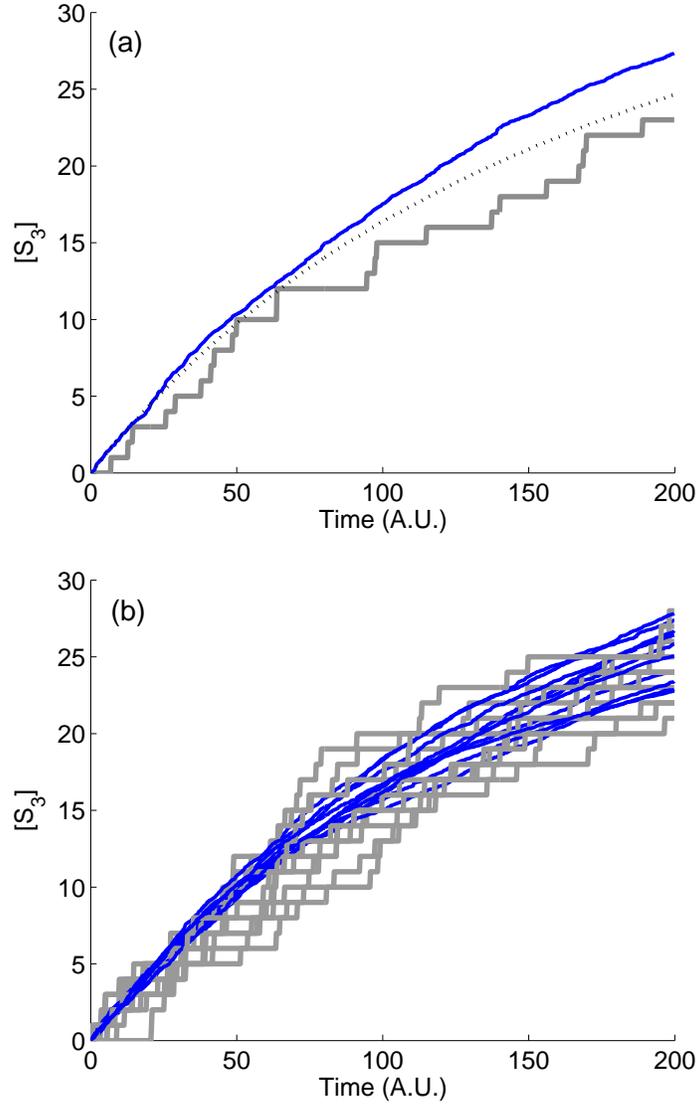}
	\caption{Dynamics of $[S_3]$  in (a) one single run and (b) 10 single runs, using Trajectory Following
	approach (gray step-like lines) and virtual population approach (dark smooth curves). The virtual population approach 
	yields smoother curves with smaller variation. The dotted line in (a) is an average over 200 runs, where both the methods are indistinguishable.}
	\label{Fig:virtualpopulation}
\end{figure}

\noindent
In Fig. \ref{Fig:virtualpopulation} (a) we present the population size of $S_3$ in a single run
 as calculated by the Trajectory Following approach and by 
the aforementioned method.
Since the population size is updated at any step, and not only in the rare events of real dimer production, the
curve is much smoother and not fluctuating. The variation between different runs in this method is smaller as well, as shown in Fig.  
\ref{Fig:virtualpopulation} (b).

\section{discussion and summary}
\label{Sec:summary}

We have proposed  a new method of analysis for stochastic  simulations data. 
The ``{\em All Possible Steps}'' method gives a precise estimation of 
 moments of the variables, which is  based on all possible steps
of the simulation. Since the information about those steps and their probabilities is calculated in any case 
throughout the simulation, the {\em All Possible Steps}  method does not require many additional computation resources. On the 
other hand, reduction in fluctuation size and variation enables one to achieve reliable statistics with
fewer runs of the algorithm. The method is applicable mainly 
for simulations under steady state condition, but it can be extended also
to some transient cases.

\begin{figure}
	\centering
		\includegraphics{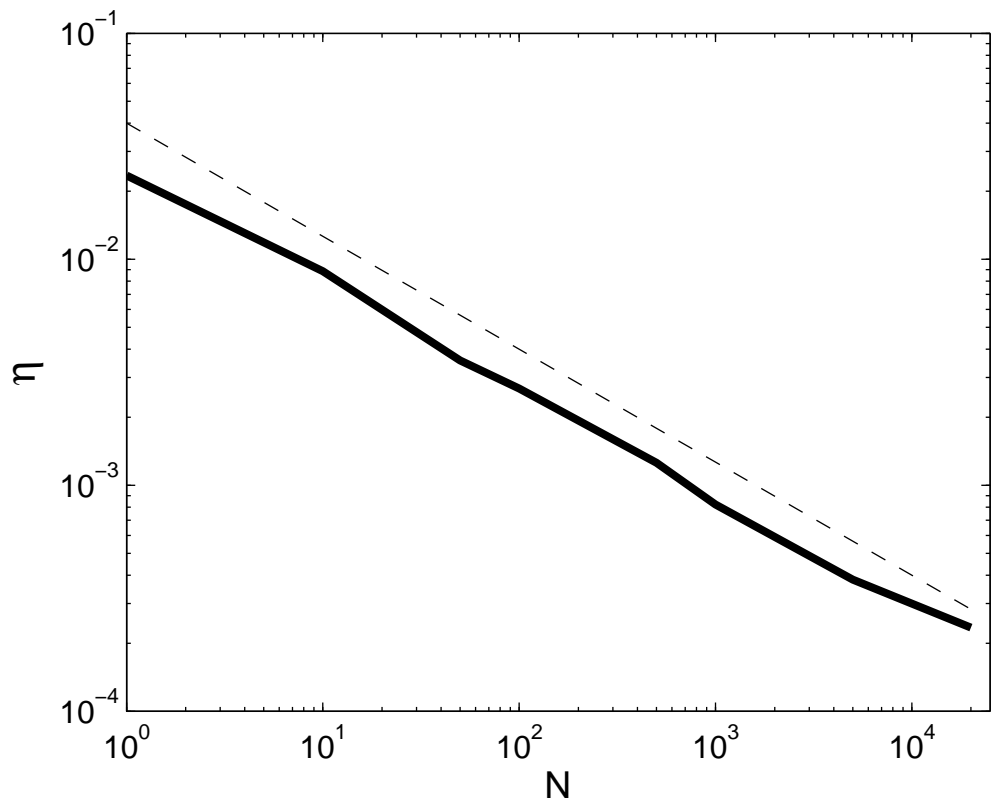}
	\caption{Estimation for the error as function of number of simulations. The dashed line is 
	proportional to $1/\sqrt{N}$. All simulation parameters are as in Fig. \ref{Fig:dimers1}.}
	\label{Fig:error}
\end{figure}

Furthermore, the availability of  two methods for moment calculation 
can be used to determine the number of runs required for any
desired level of precision. In most cases,
 there is no way of estimating how many runs are needed to obtain a given
precision. As a result, in the typical case a large number of simulations are being performed,
many more than really required. 
The existence of two estimation methods for the moments provides  a tool for the estimation of statistical error.
The relative difference between the methods is given by
\begin{equation}
\eta={|M_{\rm TF}-M_{\rm APS}| \over M_{\rm APS}},
\end{equation}
\noindent
where $ M_{\rm TF}$ and $M_{\rm APS}$ are estimations for the moment as calculated by the Trajectory Following and APS methods,
respectively. $\eta$ is a good estimator for the average error. 
Instead of defining in advance the number of simulation runs, one can run the simulation until
$\eta$ is smaller than a pre-defined threshold. This way, we avoid unnecessary simulations.
In Fig. \ref{Fig:error}, we show that $\eta$ decreases like $1/\sqrt{N}$, where $N$ is number of runs, as expected for 
the statistical error.

There are acceleration methods which are based on approximation to the exact 
Trajectory Following algorithm. This is not the case in the APS approach, which
is mathematically equivalent to averaging over many standard trajectories.  
In the Trajectory Following algorithm, the probability of a certain 
state is calculated directly as the fraction of time that  the system spent in that state.
In the {\em All Possible Steps} approach we manipulate the data obtained from the actual trajectory to reflect
virtual averaging over some hypothetical trajectories as well. 
However, we consider only the first step of each of those trajectories.
Since the actual trajectory is determined by the original algorithm, 
the number of times a state 
${\bf x}$ is being visited is not affected by considering the hypothetical steps.
As a results, the number of occurrences of a reaction ${\bf x}\rightarrow {\bf x}+{\bf \nu}$
will be calculated properly, and so is the time of staying in the new state ${\bf x}+{\bf \nu}$.

It may be possible to extend the method to include more than one hypothetical reaction.
In that case one should consider not only all possible reactions   ${\bf x}\rightarrow {\bf x}+{\bf \nu}$,
but also all possible paths ${\bf x}\rightarrow {\bf x}+{\bf \nu}_1\rightarrow {\bf x}+{\bf \nu}_2$. 
However, this requires more calculations, especially if the number of reactions is large.


The approach described here does not produce trajectories which are different from those resulting from the standard algorithms. 
It does not affect the stochastic simulation algorithm itself, but only the way  data are collected. 
 Thus, it can  apply to many  variants
of the simulation algorithm. The improvement achieved by using this method comes
{\em in addition to}, and not instead of, other algorithm-based accelerations.
Furthermore, incorporating the APS approach into some of the accelerating algorithms may enhance 
the effect of using the APS method, due to considering possible trajectories rather than possible single steps.
We leave this for future research.    
It is expected that more sophisticated versions will be developed. Progress in the two  
complementary approaches - reducing the number of required runs on the one hand, and increased efficiency 
of every single run on the other hand - will make simulations of larger systems computationally inexpensive.

\begin{acknowledgments}
I am grateful to Ravi Iyengar (MSSM) for his help. I wish to thank also Charles Peskin (NYU),  Fernand Hayot, Eric Sobie  and Padmini Rangamani (MSSM) for careful reading of the manuscript and for important comments. 
\end{acknowledgments}


\newpage

\begin{table}
	\centering
		\begin{tabular}{l|l|l|}
		& \multicolumn {1}{c|}{\bf Trajectory Following }& \multicolumn {1}{c|}{\bf All Possible Steps} \\
		\hline
		Initialization before the simulation & \multicolumn {2}{c|}{ $Q({\bf x})=0$}\\
		\hline
		Initialization before every run &\multicolumn {2}{c|} {$q({\bf x})=0$}\\
		\hline
		Update (every iteration) & \multicolumn {2}{c|} {calculate $\tau$}\\
		                         & \multicolumn {2}{c|} { $t=t+\tau$ }\\
		                         		\cline{2-3}
		                         & $q({\bf x}) = q({\bf x})+\tau$ & For each reaction $R_\mu$ \\
		                         & & $\hspace{3mm}q({\bf x}+\nu_\mu)=q({\bf x}+\nu_\mu)+{a_\mu\over a_0} \tau_\mu$\\
		\cline{2-3}
		                         &\multicolumn {2}{c|}{pick reaction $R_\mu$ and update state to ${\bf x}+\nu_\mu$}\\
	  \hline
		Normalization after every run & $Q({\bf x})=Q({\bf x})+{q({\bf x})\over t_{\rm final}-t_0}$ & $Q({\bf x})=Q({\bf x})+{q({\bf x})\over \sum_{\bf x}q({\bf x})}$ \\
		\hline
		Normalization after $N$ runs &\multicolumn {2}{c|}{$P({\bf x})=Q({\bf x})/N$}\\
		\hline
				
		\end{tabular}
		\caption{Comparison between the standard method for probability calculation and the {\em All Possible Steps} method for steady state simulations. 
		$q({\bf x})$ and $Q({\bf x})$ are probability distributions before normalization.}
			\label{Tab:compare_ss}
\end{table}

\clearpage
   
\begin{table}
	\centering
		\begin{tabular}{l|l|l|}
		& \multicolumn {1}{c|}{\bf Trajectory Following }& \multicolumn {1}{c|}{\bf All Possible Steps} \\
		\hline
		Initialization before first run & \multicolumn {2}{c|} {for each time point $t_i$, $q({\bf x},t_i)=0$}\\
		\hline
		Update (every iteration) & \multicolumn {2}{c|} {calculate $\tau$}\\
		\cline{2-3}
		                         & For each $t_i$ s.t.  $t<t_i\le t+\tau$ & $t=t+\tau$ \\
		                         & \hspace{3mm}$q({\bf x},t_i)=q({\bf x},t_i)+1$ & For each reaction $R_\mu$  \\
		                         & $t=t+\tau$ & \hspace{3mm}For each $t_i$ such that  $t<t_i\le t+\tau_\mu$ \\
		                         & &\hspace{3mm}\hspace{3mm}$q({\bf x},t_i)=q({\bf x},t_i)+a_\mu/a_0$  \\
		\cline{2-3}
		                         &\multicolumn {2}{c|}{Pick reaction $R_\mu$ and update state to ${\bf x}+\nu_\mu$}\\
		\hline
		Normalization after $N$ runs &$P({\bf x},t_i)={q({\bf x},t_i)\over N}$ &  $P({\bf x},t_i)={q({\bf x},t_i)\over \sum_{\bf x}q({\bf x},t_i)}$\\
		\hline
		\end{tabular}
		\caption{Comparison between the standard method for probability calculation and the APS method for time course simulations. 
		$q({\bf x})$ is the probability  distribution before normalization.}
			\label{Tab:compare_tc}
\end{table}

\clearpage

\newpage

\newpage

\newpage

\end{document}